\documentclass[runningheads]{llncs}
\usepackage{graphicx}
\usepackage{enumitem}
\usepackage{booktabs}
\newcommand{\pageenlarge}[1]{\enlargethispage{#1\baselineskip}}

\begin{document}
\title{KEIR @ ECIR 2025: The Second Workshop on Knowledge-Enhanced Information Retrieval}
\titlerunning{KEIR @ ECIR 2025}
%
\author{Zihan Wang\inst{1} \and Jinyuan Fang\inst{2} \and Giacomo Frisoni\inst{3} \and Zhuyun Dai\inst{4} \and Zaiqiao Meng\inst{5}  \and Gianluca Moro\inst{6} \and Emine Yilmaz\inst{7}}
\authorrunning{Z. Wang et al.}
%
\institute{
University of Amsterdam, The Netherlands. \email{z.wang2@uva.nl} 
\and
University of Glasgow, UK. \email{j.fang.2@research.gla.ac.uk}
\and
University of Bologna, Italy. \email{giacomo.frisoni@unibo.it}
\and
Google DeepMind, USA. \email{zhuyundai@google.com}
\and
University of Glasgow, UK. \email{Zaiqiao.Meng@glasgow.ac.uk} 
\and
University of Bologna, Italy. \email{Gianluca.Moro@unibo.it} 
\and
University College London, UK. \email{Emine.Yilmaz@ucl.ac.uk}
}
\maketitle              
\begin{abstract}
\vspace{-2em}
Pretrained language models (PLMs) like BERT and GPT4 have become the foundation for modern information retrieval (IR) systems. However, existing PLM-based IR models primarily rely on the knowledge learned during training for prediction, limiting their ability to access and incorporate external, up-to-date or domain-specific information. Therefore, current information retrieval systems struggle with semantic nuances, context relevance, and domain-specific issues. 
To address these issues, we propose the second Knowledge-Enhanced Information Retrieval workshop (KEIR @ ECIR 2025) as a platform to discuss innovative approaches that integrate external knowledge, aiming to enhance the effectiveness of information retrieval in a rapidly evolving technological landscape. The goal of this workshop is to bring together researchers from academia and industry to discuss various aspects of knowledge-enhanced information retrieval.

\end{abstract}
\section{Motivation}
\vspace{-0.5em}
\pageenlarge{}
In recent years, pretrained language models (PLMs)~\cite{zhu2023large} have become the foundation for modern information retrieval systems. By leveraging the vast amounts of knowledge encoded in their parameters, these models have revolutionized tasks across diverse fields, including information retrieval~(IR) and recommendation systems (RecSys). However, despite the remarkable progress in PLMs, these models primarily rely on the information learned during training for inference, limiting their ability to access and incorporate external, up-to-date, or domain-specific knowledge. As a result, PLMs can struggle with tasks requiring real-time information, specialized domain expertise, or context that extends beyond their pretraining corpus. Indeed, existing PLM-based information retrieval systems, including IR and RecSys, often encounter challenges in addressing semantic nuances~\cite{bulathwela2023leveraging}, context relevance~\cite{zhang2023variational}, and handling domain-specific intricacies~\cite{luo2022improving}, leading to imprecise results. 
To address the limitations of PLMs in accessing up-to-date or domain-specific information, recent research has proposed incorporating external knowledge into these models for enhanced performance~\cite{gao2023retrieval}. This external knowledge can come from various sources, such as external corpora~\cite{wu2024coral}, knowledge graphs (KGs)~\cite{xu2024retrieval}, or even knowledge stored within large language models (LLMs)~\cite{xi2023towards}. These methods of augmenting PLMs have shown remarkable performance in several natural language processing (NLP) tasks such as question answering~\cite{fang2024trace} and conversation generation~\cite{zhang2023variational}. 
However, leveraging external knowledge for enhancing information retrieval systems has not been fully explored. Therefore, this gap highlights the pressing requirement for innovative approaches capable of harnessing external knowledge to enhance the performance of information retrieval systems. 

To this end, we propose this second Knowledge-Enhanced
Information Retrieval workshop scheduled for ECIR 2025
(KEIR @ ECIR 2025), as a forum for the discussion of efficient and effective approaches to explore the various knowledge-enhanced information retrieval problems. This workshop addresses the growing need for more effective ways to retrieve and utilise external information in a rapidly expanding technological landscape. 

\pageenlarge{}
\textbf{KEIR @ ECIR '24.} Last year, we organised the first KEIR workshop at the ECIR 2024 Conference, which was one of the most popular workshops and was broadly welcomed by the conference attendees. During the workshop, we encouraged discussions and explorations on the following key topics:
\vspace{-0.7em}
\begin{itemize}
    \item[$\bullet$] \textbf{Knowledge-Enhanced Retrieval Models.} This topic explores the integration of external knowledge, such as KGs, to enhance the effectiveness of retrieval models. The goal is to address key challenges in retrieval models such as passage ranking~\cite{fang2023kgpr}, query reformulation~\cite{kim2021query}, and query expansion~\cite{mackie2023re}. 
    
    \item[$\bullet$] \textbf{Knowledge-Enhanced Recommendation Models.} This topic explores leveraging external knowledge as an auxiliary source of information to enrich user and item representations. The goal is to enhance the effectiveness of recommendation systems and address the cold-start problem.  
    
    \item[$\bullet$] \textbf{Knowledge-Enhanced Pretrained Language Models.} This topic explores infusing external knowledge to PLMs. Since PLMs are commonly used as the building blocks of IR models, enhancing the PLM's capability is conducive to improve the overall effectiveness of information retrieval systems. 
    
\end{itemize}

\vspace{-0.3em}\noindent Despite last year's workshop fostering optimistic discussions around the three main topics, there is still room for improvement and further exploration. Therefore, we plan to continue the discussion around these topics this year. 
Additionally, the rapid development of LLMs and retrieval-augmentation generation (RAG) models has recently lead to significantly advancements in information retrieval systems~\cite{zhu2023large}. 
Therefore, 
this year we aim to additionally promote discussions around RAG models and knowledge-aware LLMs for IR tasks. 
We encourage discussions and debate related to these areas: 
\vspace{-0.5em}
\begin{itemize}
    \item[$\bullet$] \textbf{Knowledge-Enhanced Retrieval-Augmented Generation Models.} \\
    In this topic, we will explore integrating knowledge, such as external KGs or knowledge generated by LLMs, into RAG models to improve both efficiency and effectiveness~\cite{fang2024trace}. Discussions will focus on developing effective and efficient strategies to retrieve, filter and integrate external knowledge for generating more accurate and context-aware results. The goal of integrating external knowledge into RAG models is to enhance the models' ability to retrieve more relevant information~\cite{khattab2023dspy}, reduce noises~\cite{fang2024reano} and supporting more complex reasoning tasks such as multi-hop reasoning~\cite{trivedi2023interleaving}. 
    
    \item[$\bullet$] \textbf{Knowledge-Aware Large Language Models for IR.} 
    Recently, LLMs have been widely used to enhance various aspects of IR models~\cite{DBLP:conf/sigir/Zhai24}. 
    Despite these advances, LLMs still face challenges in effectively utilizing knowledge for knowledge-intensive IR tasks~\cite{li2023personalized,yao2023react}. 
    Recent studies have also highlighted that LLMs are prone to generating incomplete, non-factual, or illogical responses~\cite{DBLP:conf/acl/XuSIC23}, due to limited knowledge awareness during vanilla fine-tuning processes~\cite{lyu2024knowtuning}. Therefore, our workshop will explore advanced fine-tuning and optimization techniques for LLMs tailored to IR systems, aiming to enhance retrieval accuracy, scalability, and personalisation. 
\end{itemize}

\vspace{-0.3em}\noindent Aligned with these core themes and driving motivations, this workshop holds a steadfast commitment to fostering collaboration among researchers engaged in the realm of knowledge integration for IR, RecSys and NLP. This gathering will serve as a platform to not only deliberate upon the advantages and hurdles intrinsic to the development of knowledge-enhanced PLMs, IR models and RecSys models but also to facilitate in-depth discussions concerning the same.

\pageenlarge{}
\section{Workshop Scope}
\vspace{-0.5em}
The workshop will concentrate on various aspects of knowledge-enhanced information retrieval, including models, techniques, data collection, and evaluation methodologies. Topics covered will include, but are not limited to:
\vspace{-0.5em}
\begin{itemize}[label=$\bullet$]
    \item Knowledge-enhanced information retrieval models.
\item Knowledge-enhanced approaches for query processing, including query parsing, query expansion, relevance feedback, and query reformulation.
    \item Knowledge-enhanced recommendation models.
    \item Knowledge-enhanced language models for retrieval.
    \item Data augmentation for knowledge-enhanced information retrieval.
    \item Large language model enhanced information retrieval. 
    \item Data collection for knowledge-enhanced information retrieval.
    \item Knowledge-enhanced retrieval-augmented generation models.
    \item Knowledge-aware fine-tuning and optimization methods for large language models tailored to IR systems.
    \item Applications of knowledge-enhanced retrievals, such as dialogue systems, question answering, summarisation and other domain-specific applications.
    \item Evaluation methodologies for knowledge-enhanced retrieval.
    \item The interpretability and analysis of knowledge-enhanced models for IR, including potential biases, fairness and ethical considerations.
\end{itemize}

\section{Format and Planned Activities}
We are organising a half-day, in-person workshop. The workshop will include keynote talks, a mix of oral and poster presentations, and interactive panel discussions to foster participant engagement.

The workshop’s call for papers invites submissions ranging from 6 to 12 pages in length. Accepted papers will have the opportunity to be published in Springer's Lecture Notes in Computer Science (LNCS) series. Additionally, authors of relevant papers accepted to the main conference may be invited to give a brief talk during the workshop.


\section{Organisers}
\vspace{-0.5em}
The organisation team consists of active IR, NLP and RecSys researchers.

\vspace{0.3em}\noindent\textbf{Zihan Wang.} 
Zihan is a fourth-year PhD student at the University of Amsterdam (UvA). He has published over ten papers in prestigious conferences including KDD, SIGIR, CCS, and WSDM. He received the Best Student Paper Award at WSDM 2018. His current research focuses on information extraction, knowledge graph embedding, and the reliability of large language models. Zihan is also co-organizing the R$^{3}$AG workshop at SIGIR-AP 2024.

\vspace{0.3em}\noindent\textbf{Jinyuan Fang.} 
Jinyuan is a third-year PhD student at the University of Glasgow. His research interests include information retrieval, knowledge graphs, and natural language processing, with a special focus on leveraging KGs to enhance retrieval-augmented generation models. He has published several papers in some top-tier conferences and journals, including KDD, NeurIPS, ACL and TOIS. He co-organised the KEIR workshop at ECIR 2024. 

\vspace{0.3em}\noindent\textbf{Giacomo Frisoni.}
Giacomo is a Postdoctoral Researcher at the University of Bologna. His research explores the intersection of Knowledge-Enhanced NLP, LLMs, and Graph Neural Networks, focusing on combining neural language models with symbolic knowledge. He has presented prolifically at top-tier conferences and journals, including ACL, EMNLP, and AAAI, receiving two best paper awards. Giacomo has served on program committees for over 12 esteemed venues and held the position of HuggingFace Student Ambassador. He co-organised the KEIR workshop at ECIR 2024.

\vspace{0.3em}\noindent \textbf{Zhuyun Dai.} Zhuyun Dai is a Staff Research Scientist at Google DeepMind. Her research interests lie in large language models, information retrieval, and machine learning. Recently, her work has concentrated on developing generalizable and capable neural retrieval models, enhancing large language model factuality, and advancing instruction fine-tuning techniques. Zhuyun actively contributes to the organizational community of various conferences in these fields, such as the co-chair of the SIGIR workshop, area chair of  ICTIR 2024,  and program committee member for many top conferences such as NeurIPS, ICLR, EMNLP, SIGIR, and AAAI. 

\vspace{0.3em}\noindent\textbf{Zaiqiao Meng.}
Zaiqiao is a Lecturer at the Information Retrieval Group of the University of Glasgow.  His research interests include IR, NLP, KG and RecSys. He is particularly interested in combining LLMs with KGs. He previously worked as a postdoctoral researcher at the University of Cambridge and the University of Glasgow. Zaiqiao has served in many community organisation roles such as the local organisation chair of ECIR 2024, the co-chair of the KEIR workshop and the program committee member for more than 16 esteemed conferences, such as ICML, NeurIPS, ICLR, ACL and EMNLP. He served as the area chair for many NLP conferences, such as ACL 2022, NAACL 2024 and EMNLP 2023. 

\vspace{0.3em}\noindent\textbf{Gianluca Moro.}
Gianluca is an Associate Professor of the University of Bologna. His research focuses on text mining, natural language processing, data mining and data science, with particular interest in Artificial Intelligence methods based on machine learning and deep neural networks, among which generative language models, to transform data into knowledge, both in structured and unstructured data domains and big data. He develops research in national and international research projects, he co-authored over ninety publications, mainly in international journals and top congresses, among which AAAI, EMNLP, IJCAI, ECAI, ACL, COLING. He has organised workshops at prestigious international conferences and collaborates on scientific projects with research centres and public and private companies. He leads the research team in NLP at DISI in Cesena.


\vspace{0.3em}\noindent\textbf{Emine Yilmaz.}
Emine is a Professor and ELLIS Fellow at University College London, Department of Computer Science. She is also an Amazon Scholar, where she works with the Alexa team. Her research interests lie in the fields of IR and NLP. She has been awarded two best paper awards (ACM ICTIR 2022 Best Student Paper and ACM CHIIR 2017 Best Paper) and was nominated for the best paper award several times (ACM WSDM 2011, ACM SIGIR 2010 and 2009). She has held various senior roles in journals and conferences, including co-editor-in-chief of the Information Retrieval Journal, editorial board member of the AI Journal, elected executive committee member of ACM SIGIR, PC Chair for ACM CIKM 2022 Applied Track, and leadership positions for conferences such as ECIR 2020, ACM SIGIR 2018, ACM ICTIR 2017, ACM SIGIR 2023 Industry Chair, WWW 2021 Panels Chair, ACM WSDM 2017 Practice and Experience Chair, and ECIR 2017 Doctoral Consortium Chair.

\pageenlarge{}

\section{Expected Audience and Advertisement}
\vspace{-0.5em}
\pageenlarge{}

Our workshop is expected to draw a diverse group of participants, including researchers, academics, industry professionals, and students from the fields of IR, NLP, RecSys, and related areas. PhD students interested in advancing retrieval techniques through the integration of external knowledge sources and PLMs will find it particularly beneficial. Industry professionals looking to enhance their search engines or recommendation systems with domain-specific knowledge graphs will discover valuable insights and networking opportunities. Academics aiming to deepen their knowledge of knowledge-driven models and their real-world applications will also find the discussions enriching.

To ensure broad visibility and engagement, the workshop will be promoted extensively through academic and research networks, professional organizations, and relevant conferences in the field. We will leverage online platforms, mailing lists, social media channels, and academic communities to reach a global audience. Furthermore, partnerships with related workshops, conferences, and research groups will enhance the workshop's visibility and impact.

\pageenlarge{}

\bibliographystyle{splncs04}
\bibliography{reference}
\pageenlarge{}
\end{document}